\documentclass[journal]{IEEEtran}
\usepackage{graphicx}
\usepackage{multirow}
\usepackage{url}
\usepackage{amsmath}
\usepackage{algorithm}
\usepackage{algpseudocode}

\begin{document}
\title{ML-Powered FPGA-based Real-Time Quantum State
Discrimination Enabling Mid-circuit Measurements}

\author{%
  \IEEEauthorblockN{%
    Neel R. Vora\textsuperscript{1},
    Yilun Xu\textsuperscript{1,*},
    Akel Hashim\textsuperscript{2},
    Neelay Fruitwala\textsuperscript{1},
    Ho Nam Nguyen\textsuperscript{2},\\
    Noah Goss\textsuperscript{2},
    Haoran Liao\textsuperscript{2},
    Jan Balewski\textsuperscript{1},
    Abhi Rajagopala\textsuperscript{1},
    Kasra Nowrouzi\textsuperscript{1},
    Qing Ji\textsuperscript{1},\\
    K. Birgitta Whaley\textsuperscript{2},
    Irfan Siddiqi\textsuperscript{2},
    Phuc Nguyen\textsuperscript{3,*},
    Gang Huang\textsuperscript{1}
  }
  \vspace{1ex}\\
  \IEEEauthorblockA{\textsuperscript{1}Lawrence Berkeley National Laboratory, Berkeley, CA, USA}\\
  \IEEEauthorblockA{\textsuperscript{2}University of California, Berkeley, CA, USA}\\
  \IEEEauthorblockA{\textsuperscript{3}University of Massachusetts Amherst, Amherst, MA, USA}
  \vspace{1ex}\\
  \normalsize\textsuperscript{*}Corresponding authors:
  Yilun Xu \texttt{(yilunxu@lbl.gov)},
  Phuc Nguyen \texttt{(vp.nguyen@cs.umass.edu)}
}

\maketitle

\begin{abstract}
Accurate and timely quantum state discrimination is critical for quantum computing, particularly in protocols requiring mid-circuit measurement (MCM) and conditional feed-forward. While classical state readout in transistors is near-instantaneous, identifying the state of superconducting qubits remains latency-limited and error-prone. Existing approaches rely on post-processing measurement data transferred to host computers, introducing delays on the order of milliseconds—vastly exceeding the coherence time of quantum states, which typically last only hundreds of microseconds.

To bridge this latency gap, we present an in-situ machine learning (ML) inference engine implemented on a field-programmable gate array (FPGA), enabling real-time quantum state discrimination within 40\,ns. Our system supports both two-level (qubit) and three-level (qutrit) systems and performs inference directly on digitized readout signals without host-side intervention. This low-latency operation facilitates mid-circuit measurements, where qubit states are measured and reused within a single quantum circuit, and enables real-time conditional operations essential for quantum error correction.

Crucially, fast and accurate mid-circuit readout is a prerequisite for implementing practical fault-tolerant error correction, where errors must be detected and corrected within a single circuit cycle to preserve logical state coherence.

We validate our system using superconducting transmon devices, demonstrating robust discrimination fidelity across multiple qubit and qutrit channels. Further, we implement a conditional qutrit logic protocol based on FPGA-resident state classification, highlighting the practical benefits of our approach for NISQ-era quantum algorithms and scalable fault-tolerant architectures.
\end{abstract}


\pagestyle{plain}

{\section{\textbf{Introduction}}}
\label{sec:introduction}

\begin{figure*}[t]
    \includegraphics[width=1\linewidth]{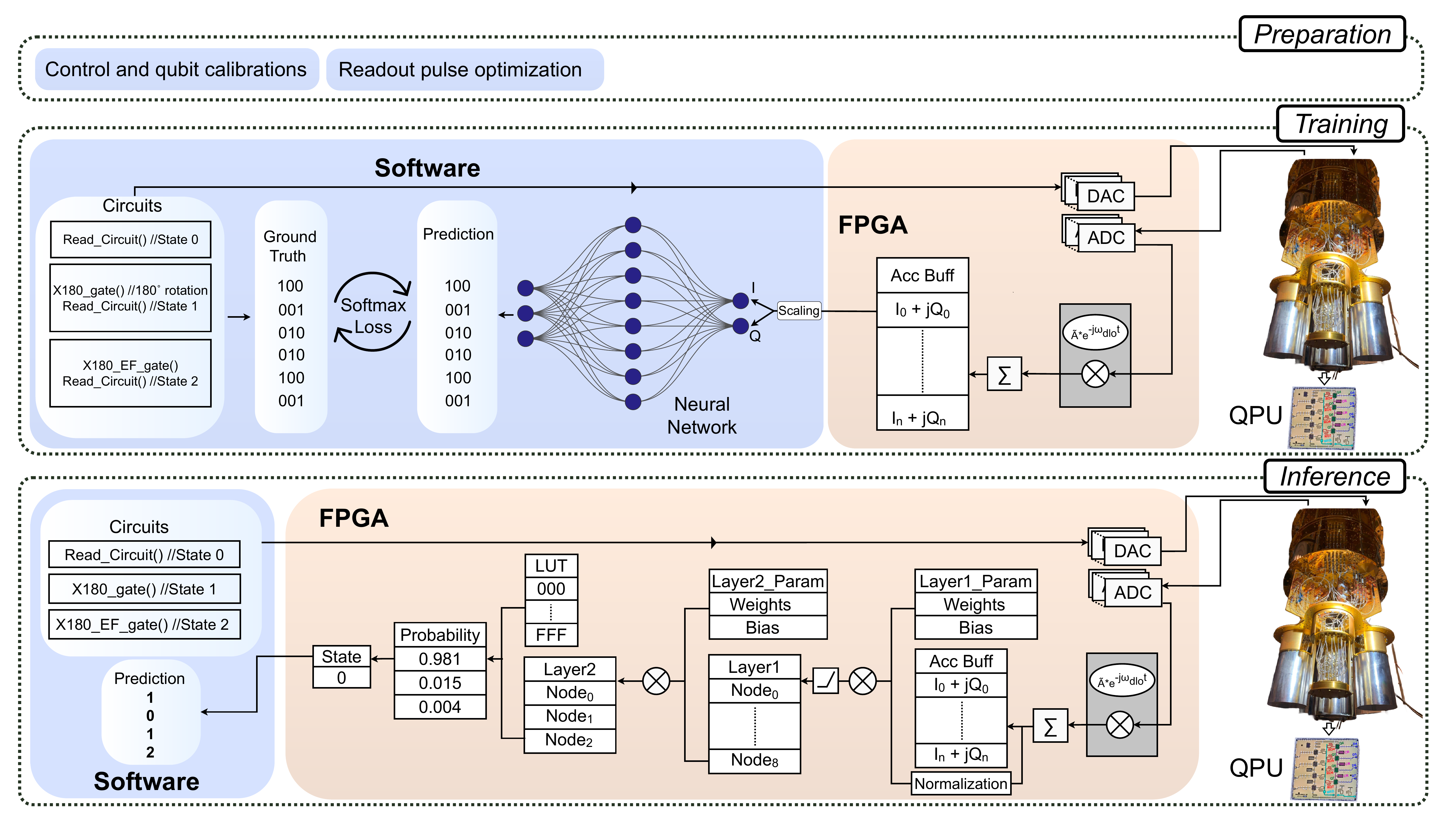}
   \caption{QubiCML system design for FPGA-enabled real-time quantum state discrimination. Top: Supervised training loop using offline IQ data and neural network optimization in software. Bottom: Inference pipeline deployed on FPGA, performing real-time state classification using quantized network weights and IQ samples directly from the quantum processor.}

    \label{fig:system_design}
\end{figure*}

Quantum computing is a rapidly evolving technology that holds the promise of revolutionizing computation across various domains thanks to its ability to solve specific classes of complex problems. Quantum computers have strong computational advantages compared to classical computers for certain problems as they leverage the quantum-mechanical properties of quantum bits (qubits). In the current practice, qubit information is manipulated by an electronic control system, which includes qubit readout and control pipelines, to orchestrate the execution of quantum programs. The control pipeline transmits precise gate pulses to qubits while the readout pipeline measures the qubits' states. Control and readout operations can be executed on hundreds of qubits~\cite{ibm127}, while a fully utilized 50-100 qubit quantum computer may be able to perform tasks that surpass the capabilities of today’s classical digital computers~\cite{preskill_quantum_2018}. 

 Among many advanced qubit architectures introduced recently~\cite{kjaergaard_superconducting_2020,bruzewicz_trapped_ion_2019}, superconducting qubits~\cite{kjaergaard_superconducting_2020} have emerged as the leading quantum computing platform in the past two decades. 
Superconducting qubit information can be obtained by state measurement~\cite{krantz2019quantum}, a process to identify the qubit state. However, this is the slowest operation in quantum computing. While individual superconducting qubit supports around 100~$\mu s$ coherence time (i.e., time to hold the information), the delay in streaming data from the readout measurement circuit to the host computer for state discrimination falls within the millisecond ($ms$) range. Hence, real-time state measurement with in-loop host computer is not possible. In the current practice, the measurement must be done at the end of the circuit, which is normally a classical host computer. The state discrimination results are computed on a classical computer~\cite{LienhardBenjamin, QSD_2010, QSD_2022}, where it is then subjected to classical post-processing.

\textbf{Mid-circuit measurement (MCM)}—the ability to \emph{selectively} collapse, conditionally reset and reuse qubits inside a computation. MCM is crucial for \textbf{quantum error correction (QEC)}. It allows for extracting information about a qubit's state during a computation, enabling real-time error detection and correction without halting the entire process. This is essential for building fault-tolerant quantum computers \cite{mcmXqce}. MCMs allow for the detection of errors as they occur during a quantum computation. This is vital because errors can propagate and amplify if not addressed promptly. MCM thus imposes stringent demands on discrimination hardware: analog IQ signals must be digitized, classified, and a decision issued well within the qubit coherence window to prevent decoherence-driven errors from propagating into subsequent circuit layers. Therefore, demands \emph{in-situ} discriminators capable of delivering a binary decision on the same RFSoC clock domain that drives the qubits, eliminating host-bound round-trips and unlocking deterministic feed-forward

Recent work has explored machine learning (ML) and hardware co-design to accelerate qubit-state discrimination. Ristè \emph{et al.} demonstrated closed-loop feedback reset using a single threshold discriminator on a non-FPGA device, with threshold optimized to give $\sim$99\% single-shot two-level readout fidelity \cite{riste2012feedback}. Salathé \emph{et al.} built an FPGA-based low-latency readout/feedback chain where state discrimination is threshold-based (no ML), achieving peak readout fidelity $\sim$77\% and $\sim$110 ns input-to-feedback latency \cite{salathe2018low}. Reuer \emph{et al.} implemented reinforcement learning based decision-making on the FPGA, reporting $\sim$98.05\% two-level readout fidelity and $\sim$88.7\% for three-level with sub-µs control latency \cite{reuer2023realizing}. Sivak \emph{et al.} realized real-time QEC control where the ancilla state discrimination on FPGA is non-ML two-line thresholding on integrated I/Q \cite{sivak2023real}. Benjamin \emph{et al.} showed that a feed-forward neural network (FNN) improves assignment fidelity for multiplexed transmon readout, but their implementation was hardware-inefficient \cite{LienhardBenjamin}. Satvik \emph{et al.} and follow-on work reduced complexity through hardware-aware designs~\cite{uwisconsin_work,10.1109/DAC63849.2025.11133314}, while Gautam \emph{et al.} explored low-latency SVM-based discrimination~\cite{Gautam2024LowlatencyML}; however, these approaches did not demonstrate end-to-end in-situ deployment within the same low-latency control hardware while maintaining the high readout fidelity. Hashim \emph{et al.}demonstrated mid-circuit measurements for active feedback using a single non-statistical approach, highlighting the need for more sophisticated qubit discrimination methods for efficient mid-circuit measurement \cite{akel_mcm}. Baumer \emph{et al.} proposed a dynamic decoupling scheme for mid-circuit measurement, but their approach required longer readout times of 1.2 $\mu s$ with a feed-forward time of ~650 $ns$ \cite{ibm_mcm}. More recent efforts have mapped neural networks to RFSoC using HLS~\cite{di2025end}, explored LUT-based classifiers~\cite{farooq2025luna}, and investigated AI Engine-based inference~\cite{Butko:2026pwb}, further expanding the design space for quantum readout accelerators. Despite these advances, no prior study demonstrates an ML discriminator that \emph{completes the entire MCM loop}—discrimination, conditional branching in a real-time setting.

To address this gap, we integrate data-processing logic directly onto the FPGA fabric (Fig.~\ref{fig:system_design}). Our system deploys a quantized neural network that completes inference in \(40\)~ns, supporting both qubit and qutrit mid-circuit measurements. We validate across multiple superconducting qubits~\cite{aqt}, demonstrating real-time MCM on two- and three-level systems. \textbf{This is the first end-to-end low-latency ML state-discrimination engine on RFSoC for MCM} \cite{vora2024ml}, closing the latency frontier for scalable quantum error correction. Our architecture is fully parallelizable across channels, scales efficiently with qubit number, and supports dynamic branching, making it compatible with modular quantum control stacks. By coupling near-optimal fidelity with ultralow latency, this platform enables deterministic feedback, qubit reuse, and conditional circuit evolution—all essential for fault-tolerant operation in both NISQ and post-NISQ regimes.

This integration of ML-based state discrimination into FPGA-based control hardware represents a significant advancement toward real-time, high-fidelity qubit measurement, which is essential for the scalability and practical implementation of quantum computing systems.

\vspace{5pt}
\section{\textbf{Result}}
\label{sec:result}

\begin{figure*}[t]
    \centering
    \includegraphics[width=1\linewidth]{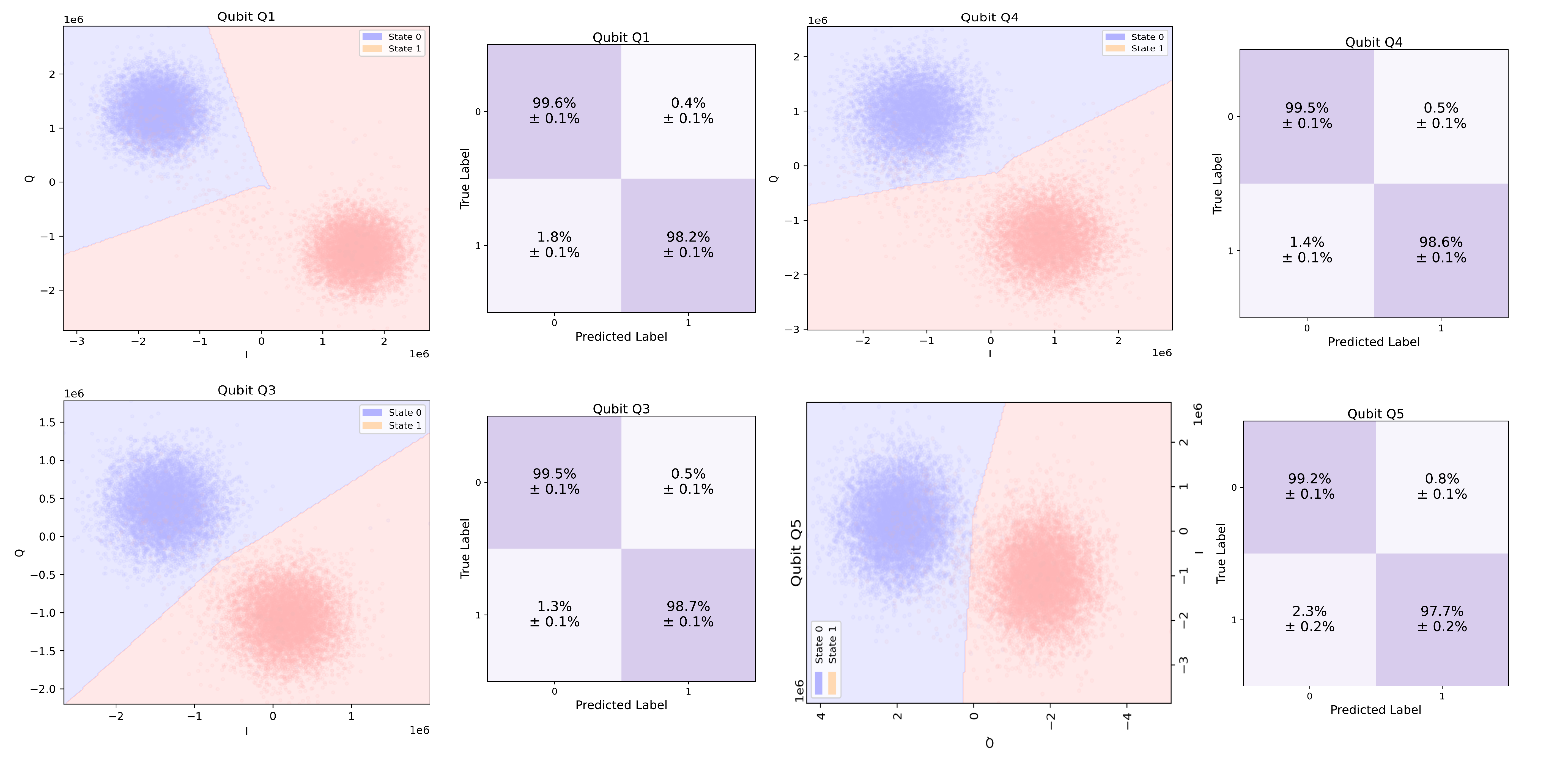}
    \caption{Real-time FPGA-based state discrimination for two-level quantum systems. Each cluster corresponds to measurement outcomes in the IQ plane for $|g\rangle$ and $|e\rangle$ states.}
    \label{fig:Qubit_res}
\end{figure*}  

Rapid and accurate qubit state measurement is crucial for effective quantum error correction and the overall performance of quantum computing systems. In this section, we demonstrate the system's capability to distinguish between the ground and first excited states of a qubit with high fidelity and low latency. Extending the discrimination to include higher energy states, we showcase the system's effectiveness in identifying qutrit states, which is essential for advanced quantum computing applications. The system's low-latency processing enables mid-circuit measurements, allowing for the implementation of more complex quantum algorithms. Leveraging real-time state discrimination.

These capabilities are achieved by deploying a neural network architecture onto an FPGA-based system. The FPGA hosts the base architecture of the neural network, while the specific weights for each qubit are trained offline and subsequently loaded onto the FPGA. This integration enables state discrimination within 40ns (without softmax), facilitating state-based decision branching.

\subsection{\textbf{Two-Level Quantum State Discrimination}}
We demonstrate real-time quantum state discrimination for two-level systems (\textit{i.e.}, qubits) using a low-latency FPGA-based architecture. Quantum circuits were compiled via the QubiC software stack, incorporating calibrated single-qubit $\pi$-pulses with tailored parameters—frequency, phase, amplitude, and DRAG coefficients—for each of the eight transmon qubits under test. Readout pulses were engineered with cosine-edge envelopes and configured to match the resonance frequencies of their respective readout resonators. A digital local oscillator (DLO), phase-locked to the readout frequency, was synthesized on-chip to enable homodyne mixing with the readout signal.

Once the qubits were driven into their excited states, all eight were read out simultaneously. The analog return signal, comprising superimposed responses from each qubit, was digitized by the FPGA’s ADCs. Digitized readout was then digitally mixed with its respective DLO to extract individual readouts. These were integrated in time to yield a single complex point in the in-phase/quadrature (IQ) plane per qubit.

To classify the quantum states, we collected approximately 5,000 labeled readout shots each for the ground ($|g\rangle$) and excited ($|e\rangle$) states. A multilayer perceptron (MLP) was trained offline and subsequently deployed to the FPGA. The same compact MLP architecture is reused for all qubits, with one instance per channel running inference in parallel. During initialization, the quantized model weights were loaded into dedicated memory blocks, and during execution, the integrated IQ points were streamed into the on-chip MLP pipelines for low-latency classification.

As shown in Fig.\ref{fig:Qubit_res}, the system accurately discriminated the $|g\rangle$ and $|e\rangle$ states across four representative qubits. Complete fidelity metrics for all eight are summarized in Table\ref{tab:Qubit_res}. The entire pipeline—from readout acquisition to state prediction—was executed within 40,ns (or 66,ns with Softmax), demonstrating compatibility with real-time quantum feedback constraints.

\begin{table}[t]
    \centering
    \begin{tabular}{l||c|c|c}
    \hline
      Qubit& $0|0\rangle$ &$1|1\rangle$ & Avg\\
     \hline
      \hline
        Q0 (500ns) & 98.58 ± 0.12\%   & 97.72 ± 0.16\%  &98.15 ± 0.10\%\\
        Q1 (650ns) & 99.73 ± 0.05\%   & 98.55 ± 0.12\%  &99.14 ± 0.07\%\\
        Q2 (700ns) & 99.01 ± 0.10\%   & 90.62 ± 0.30\%  &94.82 ± 0.16\%\\
        Q3 (650ns) & 99.66 ± 0.06\%   & 98.16 ± 0.14\%  &98.91 ± 0.08\%\\
        Q4 (750ns) & 99.48 ± 0.07\%   & 98.47 ± 0.12\%  &98.98 ± 0.07\%\\
        Q5 (750ns) & 99.34 ± 0.08\%   & 97.57 ± 0.16\%  &98.45 ± 0.09\%\\
        Q6 (700ns) & 98.46 ± 0.13\%   & 96.99 ± 0.18\%  &97.72 ± 0.11\%\\
        Q7 (500ns) & 99.78 ± 0.04\%   & 96.18 ± 0.19\%  &97.98 ± 0.10\%\\
     \hline
    \end{tabular}
     \vspace{5pt}
    \caption{System performance evaluated on multiple qubits on Chip X6Y3 AQT \cite{aqt}}
    \label{tab:Qubit_res}
\end{table}

\begin{figure*}[t]
    \centering
    \includegraphics[width=1\linewidth]{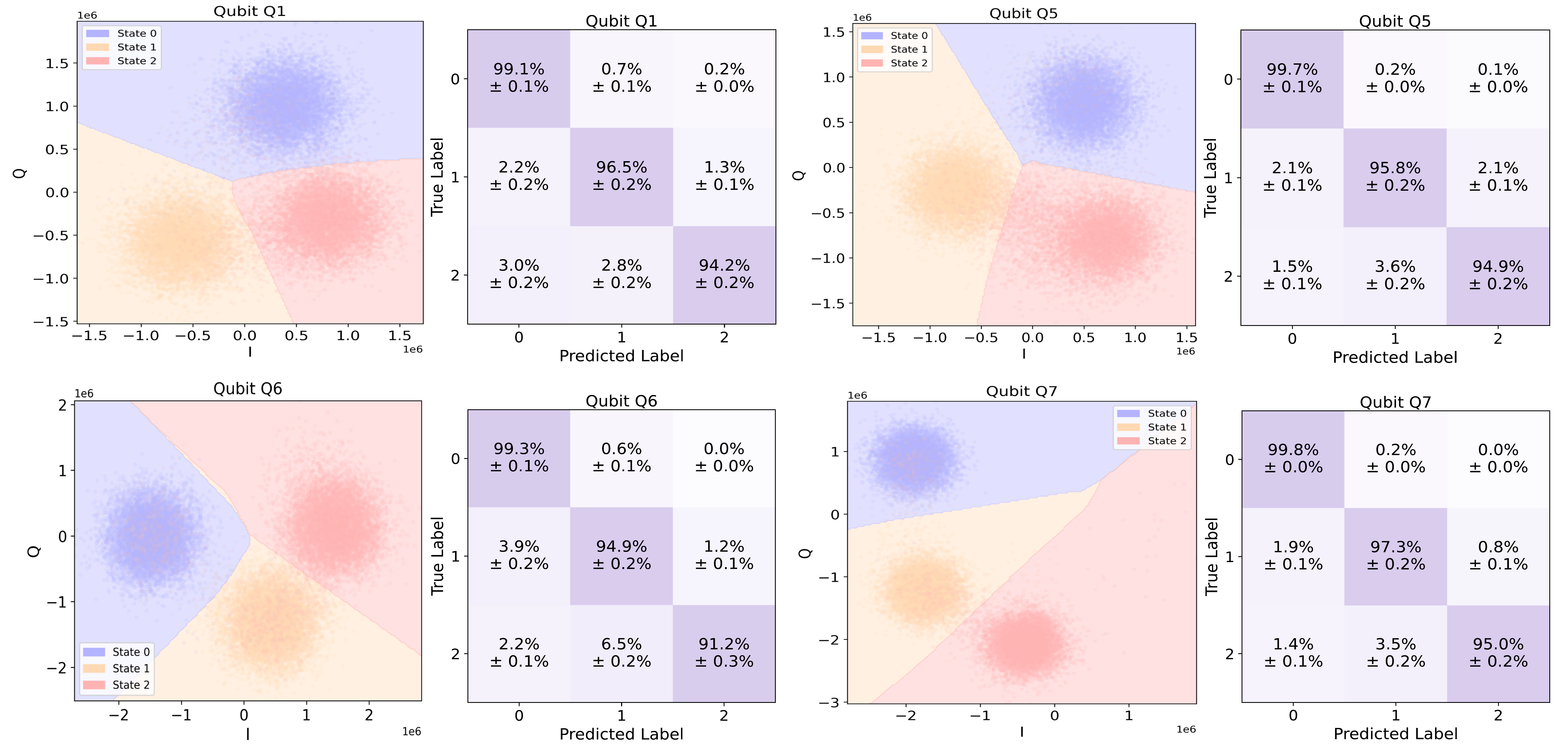}
    \caption{FPGA-based real-time state discrimination results for a three-level (qutrit) quantum system. IQ clustering shows well-separated state distributions for $|g\rangle$, $|e\rangle$, and $|f\rangle$.}
    \label{fig:Qutrit_res}
\end{figure*}

\subsection{\textbf{Three-Level Quantum State Discrimination}}
We extend our approach to three-level systems (qutrits) by incorporating the second excited state $|f\rangle$ of the transmon qubit. Due to transmon anharmonicity, the $|e\rangle \rightarrow |f\rangle$ transition occurs at a distinct frequency, allowing selective excitation via an EF $\pi$-pulse without perturbing the GE manifold.

To prepare the qutrit states, we designed circuits that initialized: (i) $|g\rangle$ via readout only, (ii) $|e\rangle$ using a single GE $\pi$-pulse, and (iii) $|f\rangle$ using a composite sequence of GE and EF $\pi$-pulses. Control pulses and readout signal paths were optimized as in the two-level case.

After excitation, all eight qubits were read out concurrently. The return signal was digitized and demodulated on the FPGA using per-qubit DLOs. Integrated IQ points for each qubit were then passed into independent MLP classifiers 
each trained on $\sim$5,000 examples per class ($|g\rangle$, $|e\rangle$, $|f\rangle$). Notably, the exact same neural network architecture used for qubit classification is reused for qutrit discrimination, highlighting its scalability across Hilbert space dimensionality.

As shown in Fig.\ref{fig:Qutrit_res}, the system reliably classified the three qutrit states within the same 40,ns inference latency. Full fidelity metrics are presented in Table\ref{tab:Qutrit_res}. To our knowledge, this marks one of the first real-time, FPGA-resident implementations of qutrit state discrimination—enabling scalable mid-circuit measurement and control in higher-dimensional quantum protocols.

\begin{table}[t]
    \centering
    \begin{tabular}{l||c|c|c|c}
    \hline
      Qubit& $0|0\rangle$ &$1|1\rangle$  &$2|2\rangle$ & Avg\\
     \hline
      \hline
        Q0 (500ns) & 99.2 ± 0.1\%   & 94.1 ± 0.2\%  &92.9 ± 0.3\% &95.4 ± 0.1\%\\
        Q1 (650ns) & 99.1 ± 0.1\%   & 96.5 ± 0.2\%  &94.2 ± 0.2\% &96.7 ± 0.1\%\\
        Q2 (500ns) & 99.1 ± 0.1\%   & 93.5 ± 0.3\%  &90.9 ± 0.3\% &94.5 ± 0.1\%\\
        Q3 (650ns) & 99.6 ± 0.1\%   & 96.8 ± 0.2\%  &88.0 ± 0.3\% &94.8 ± 0.1\%\\
        Q4 (750ns) & 99.7 ± 0.1\%   & 94.3 ± 0.2\%  &90.2 ± 0.3\% &94.7 ± 0.1\%\\
        Q5 (750ns) & 99.8 ± 0.1\%   & 95.9 ± 0.2\%  &94.9 ± 0.2\% &96.8 ± 0.1\%\\
        Q6 (700ns) & 99.4 ± 0.1\%   & 94.9 ± 0.2\%  &91.2 ± 0.3\% &95.2 ± 0.1\%\\
        Q7 (500ns) & 99.8 ± 0.1\%   & 97.3 ± 0.2\%  &95.0 ± 0.2\% &97.3 ± 0.1\%\\
     \hline
    \end{tabular}
    \vspace{5pt}
    \caption{System's performance evaluated on multiple qubits on Chip X6Y3 AQT \cite{aqt}}
    \label{tab:Qutrit_res}
\end{table}

\begin{figure}[h]
    \includegraphics[width=1\linewidth]{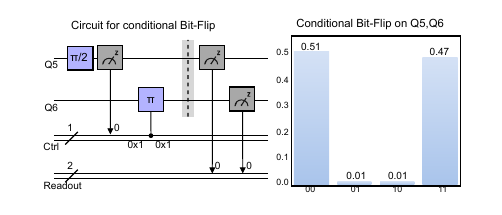}
    \vspace{-10pt}
    \caption{\textbf{Conditional bit-flip using mid-circuit measurement on qubits Q5 and Q6.} Circuit diagram (left) illustrating initial state preparation, mid-circuit measurement via FPGA-based discrimination, and conditional gate execution. Resulting state populations (right) confirm effective conditional logic implementation.}
    \label{fig: mid_circuit}
\end{figure}

\begin{figure*}[t]
    \centering
    \includegraphics[width=1\linewidth]{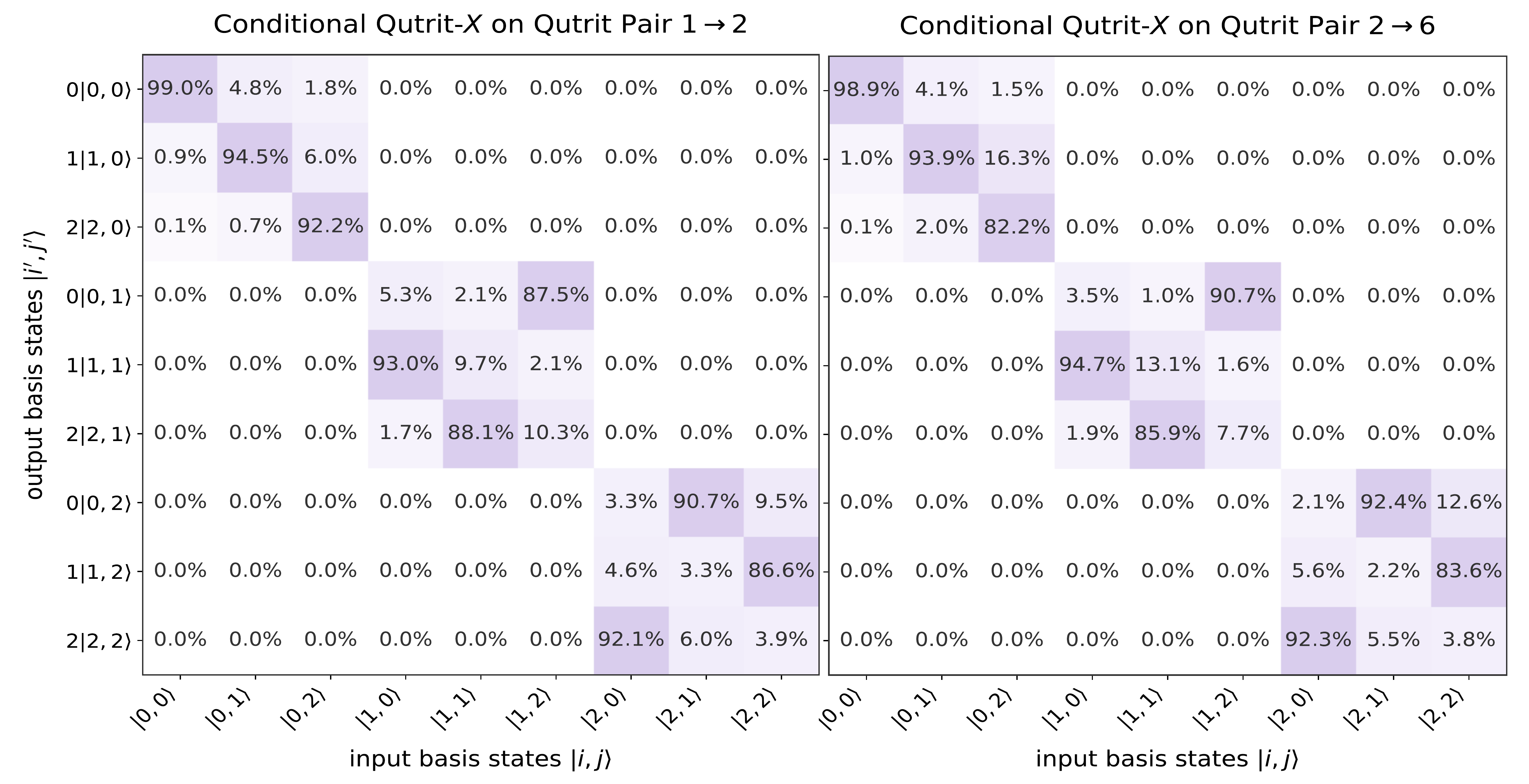}
    \caption{\textbf{Confusion matrices for conditional qutrit-\(X\) operations based on mid-circuit measurement outcomes.} 
    We implement conditional single-qutrit operations on two control-target qutrit pairs \({1 \rightarrow 2}\) (left) and \({2 \rightarrow 6}\) (right) inspired by the three-level conditional-SUM (CSUM) transformation. Each matrix shows the measured output populations following preparation of the nine input basis states \(|i, j\rangle\), mid-circuit measurement of the control qutrit, on-chip neural inference via FPGA, and conditional qutrit-\(X\) gates applied to the target. The measured transitions reflect modular arithmetic behavior \(|i, j\rangle \rightarrow |i, i{+}j \bmod 3\rangle\), validating real-time conditional logic. Notably, these results are shown \textit{without} readout correction;}
    \label{fig:CSUM}
\end{figure*}

\subsection{\textbf{Mid-Circuit Measurements with Conditional Bit-Flip}}

Mid-circuit measurement (MCM) is fundamental for enabling conditional logic and real-time feedback in quantum circuits. To illustrate our FPGA-based real-time discrimination and feed-forward capability, we implemented a two-level conditional bit-flip protocol on transmon qubits Q5 and Q6. The experimental procedure, depicted in Fig.~\ref{fig: mid_circuit}, begins by preparing Q5 on the equator (Super-position)via a calibrated \(\pi/2\)-pulse, followed by an immediate mid-circuit measurement performed through QubiCML’s integrated FPGA-based neural inference engine. The measurement outcome from Q5 directly conditions the subsequent operation on Q6: if Q5 measures in the \(|1\rangle\) state, a conditional bit-flip (X gates) is executed on Q6; if Q5 measures \(|0\rangle\), no operation is applied. The final distribution, exhibiting dominant population peaks at \(|00\rangle\) and \(|11\rangle\), confirms successful execution of conditional logic via mid-circuit measurement and immediate feed-forward. This two-level demonstration sets the stage for more complex multi-level operations, such as the three-level CSUM protocol discussed subsequently.

\subsection{\textbf{Conditional Qutrit‑X with Mid‑Circuit Discrimination}}

To demonstrate the utility of real‑time FPGA‑resident state discrimination in mid‑circuit protocols, we implemented a conditional qutrit‑\(X\) operation based on measurement of a control qutrit, serving as a precursor to a full three‑level CSUM gate \cite{CSUM}. This setup benchmarks the integrated discrimination and conditional feed‑forward pipeline in a multi‑level quantum system.

While the ideal CSUM performs \( |i,j\rangle \rightarrow |i, (i+j)\bmod 3\rangle\), our current experiment applies a conditional single‑qutrit operation on the target, focusing on mid‑circuit discrimination under realistic constraints.

\paragraph{Experimental protocol:}
\begin{enumerate}
  \item \textbf{State preparation:} All nine basis states \(|i,j\rangle\) are initialized using sequential pulses driving the \(g \to e\) and \(e \to f\) transitions.
  \item \textbf{Control measurement:} Perform a mid‑circuit measurement of the control qutrit using the FPGA‑based MLP discriminator.
  \item \textbf{Conditional qutrit‑\(X\):} Based on the measurement outcome \(m\):
  \begin{itemize}
    \item If \(m=0\), no operation is applied.
    \item If \(m=1\), apply a single qutrit‑\(X\): first an \(e \to f\) pulse, then a \(g \to e\) pulse.
    \item If \(m=2\), apply two sequential qutrit‑\(X\) operations (i.e., an \(X^2\) gate).
  \end{itemize}
\end{enumerate}

We validated this protocol on two control–target qutrit pairs (\({2 \rightarrow 6}\) and \({1 \rightarrow 2}\)), measuring the truth table across all \(|i,j\rangle\) inputs. Figure~\ref{fig:CSUM} shows output distributions that qualitatively follow the modular‑arithmetic pattern expected for a full CSUM operation, even though only conditional qutrit‑\(X\) operations were employed.

Figure~\ref{fig:CSUM} presents a confusion matrix using \emph{raw} measurement values—i.e., without any readout correction. The \(|0,0\rangle \rightarrow |0,0\rangle\) branch yielded fidelities of \(98.9\%\) and \(99.0\%\) for the two device pairs, corresponding to the \(m=0\) case where no conditional gate was applied. In contrast, output states in the \(m=2\) branch showed slightly lower fidelity, such as \(|2,2\rangle \rightarrow |2,1\rangle\) with \(83.6\%\) and \(86.6\%\), and \(|2,1\rangle \rightarrow |2,0\rangle\) with \(85.9\%\) and \(88.1\%\). These representative values highlight conditional logic across ternary states using real-time mid-circuit measurement and feed-forward control.

These results collectively establish the effectiveness of our FPGA-integrated real-time state discrimination framework for executing high-fidelity conditional logic in three-level quantum systems. Our architecture thus sets a critical performance benchmark and paves the way for scalable qutrit-based quantum computing applications requiring rapid conditional feedback and mid-circuit quantum control.

\subsection{\textbf{Resource Usage}}
Assessing resource utilization is crucial to ensure the efficient operation of the our system on FPGA. For a single machine learning inference pipeline deployed on the AMD Xilinx ZCU216 RFSoC, the implementation consumes only 1529 lookup tables (LUTs), 2597 flip-flops (FFs), 1 BRAM block, 43 DSP slices, and 86 CARRY8 primitives. These correspond to just 0.37\% of available LUTs, 0.30\% of FFs, 0.09\% of BRAM, and 0.90\% of DSPs on the device. 

\begin{figure}[h]
    \includegraphics[width=0.8\linewidth,trim=0 0 0 0,clip]{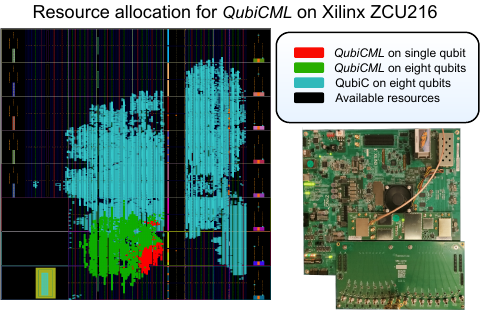}
    \vspace{-10pt}
    \caption{Resource allocation}
    \vspace{-10pt}
    \label{fig:resource}
\end{figure}

This low footprint enables parallel deployment of multiple discrimination pipelines—one per qubit—without exceeding resource constraints. Figure~\ref{fig:resource} visualizes this allocation: red indicates memory usage for a single ML pipeline, green shows cumulative usage across 8 pipelines, and blue/cyan represents memory reserved by the QubiC system. This efficiency confirms the scalability of our architecture for low-latency, multi-qubit quantum control.


\section{\textbf{Discussion}}
\label{sec:discussion}

Fast and accurate quantum state discrimination is essential for enabling mid-circuit measurements (MCMs), a critical component of quantum error correction and conditional logic. Historically, host-side post-processing has imposed latency bottlenecks, hindering real-time feedback. Our work eliminates this gap by embedding a machine learning (ML) inference engine directly within the RFSoC FPGA, achieving sub-40 ns latency from digitization to classification. This allows feed-forward operations conditioned on measurement outcomes to occur within the circuit coherence time.

The in-situ inference model generalizes beyond qubits to three-level systems (qutrits), demonstrating robust classification of both binary and ternary states. This versatility supports a broader class of quantum protocols, including those exploiting higher-dimensional Hilbert spaces for enhanced encoding and error resilience.

Relative to traditional techniques such as thresholding or matched filters~\cite{boxcar1, boxcar2, matched1, matched2}, our ML-driven approach achieves comparable fidelity while offering tunability, robustness to drift, and hardware efficiency through quantization. By operating entirely on-chip, we eliminate PCIe or host-CPU latency, unlocking new modes of conditional control within a single experimental cycle.

Crucially, our approach supports real-time feedback required by fault-tolerant quantum error correction codes, enabling fast syndrome extraction and dynamic circuit adaptation. In the NISQ era, this technology further empowers adaptive algorithms such as variational ansätze and phase estimation, where decisions based on intermediate measurements are essential.

In conclusion, this work presents the first end-to-end ML-based state discrimination system embedded within RFSoC for real-time mid-circuit measurements. By bridging the latency gap and supporting both binary and ternary state classification, we establish a scalable, reconfigurable platform for conditional logic in quantum computing, setting the stage for future error-corrected systems.

While the current system ingests integrated \(I\!Q\) data, we aim to enhance readout fidelity by deploying advanced, hardware‑efficient signal‑processing filters directly on the FPGA. These will improve SNR by removing baseline noise and crosstalk before \cite{pathsig} classification and enable time‑series‑based inference to detect and correct for \(T_1\)‑induced decay during the measurement window. By analyzing \(I(t),Q(t)\) trajectories rather than only average values, the classifier can identify mid‑pulse relaxation events, further boosting discrimination accuracy in both qubit and qutrit readout.


\section{\textbf{Method}}
\label{sec:method}

\begin{figure}
    \centering
        \includegraphics[width=0.96\linewidth, trim = 1 1 1 1, clip]{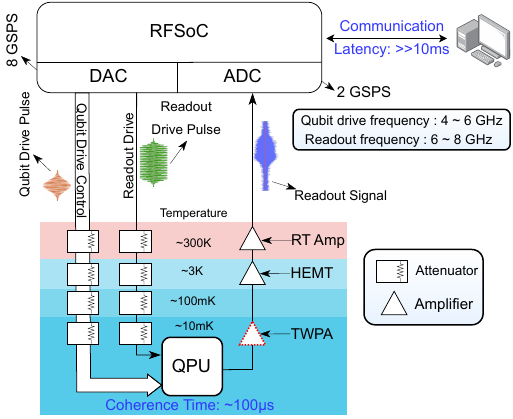}
        \vspace{-8pt}
        \caption{Schematic of common superconducting transmon qubit control and readout.}
        \vspace{-10pt}
        \label{fig:readout_architecture}
\end{figure}

\begin{figure*}
    \includegraphics[width=1\linewidth]{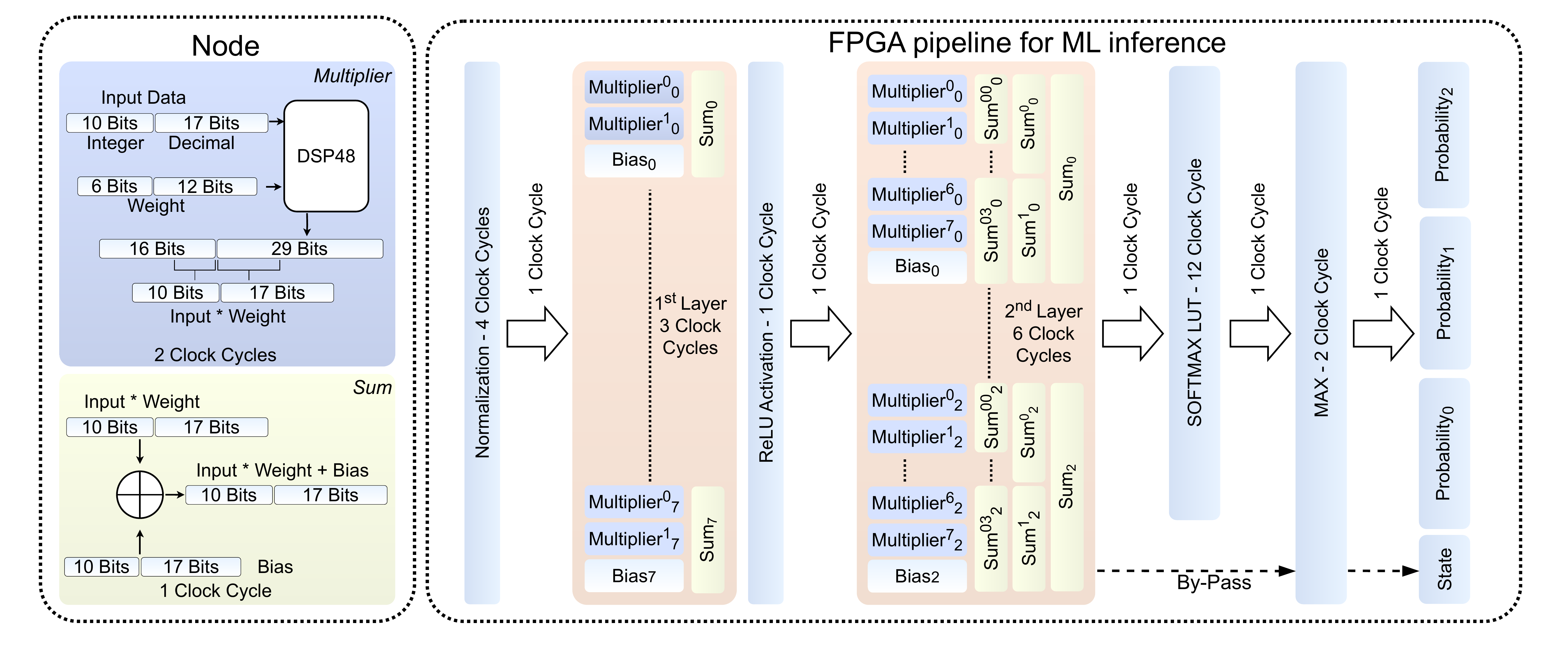}
    \caption{Pipelined fixed-point neural network implementation on FPGA. The design leverages DSP48-based multipliers, layer-wise accumulation, and quantized weights and biases to enable sub-40 ns inference latency. The softmax and max operations are optimized using LUT-based logic for single-shot classification.}

    \label{fig:fpga_implementation}
\end{figure*}

\begin{figure}[t]
    \includegraphics[trim=0 5 0 0,clip]{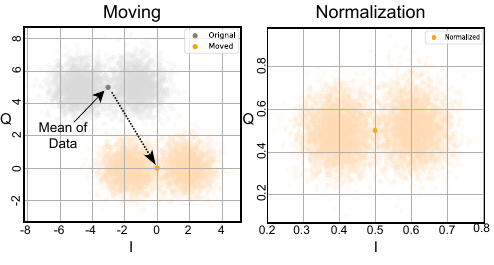}
    \caption{Data scaling using mean shifting and power-of-two normalization.}
    \label{norm_mov}
\end{figure}

\subsection{\textbf{Quantum processor and readout architecture}}

Our system is built upon the flexible QubiC 2.0 RFSoC platform~\cite{QubiC,xu2023qubic}. As illustrated in Fig.~\ref{fig:readout_architecture}, the RFSoC’s DACs generate user-defined drive pulses, which are routed through the dilution refrigerator to the quantum processing unit (QPU) operating at \(\sim 10\,\mathrm{mK}\). Readout pulses are similarly generated and directed to the QPU, stimulating the readout resonator.

The resulting response signal is routed back through the cryogenic amplification chain—passing through the traveling wave parametric amplifier (TWPA) ~\cite{doi:10.1126/science.aaa8525}, high-electron-mobility transistor (HEMT) amplifier, and room-temperature amplifier—before reaching the ADCs on the RFSoC. On the FPGA fabric, a digital downconversion (DDC) stage processes the acquired waveform, followed by a quantized neural network that performs in-situ state discrimination within 40\,ns. This tight integration eliminates the latency overhead of transferring measurement data to a host CPU.

Compared to commercial systems such as Zurich Instruments~\cite{zurich_readout} and Keysight~\cite{keysight_readout}, which are closed and difficult to modify, QubiC 2.0 offers full-stack accessibility—from gateware and firmware to software APIs—making it uniquely suitable for hardware-software co-design, including ML-accelerated quantum control and mid-circuit measurement pipelines.


\subsection{\textbf{Neural network model}}
\label{subsec: nn_model}

We implement a lightweight feedforward neural network (FNN) \cite{FNN} designed for FPGA compatibility and optimized for real-time inference. The architecture consists of:

\begin{itemize}
    \item An input layer with 2 neurons corresponding to the scaled In-phase (I) and Quadrature (Q) components.
    \item A hidden layer with 8 neurons and ReLU activation \cite{relu}.
    \item An output layer with 3 neurons followed by a softmax activation for multi-class qubit state discrimination \cite{NIPS1989_0336dcba}.
\end{itemize}

The network is trained using a dataset comprising 10,000 readout trajectories (shots) for each qubit, covering all state classes. We employ the Adam optimizer with binary cross-entropy loss, and training is performed using 32-bit floating point precision. The dataset is split into training (60\%) and validation (40\%) sets. 

To ensure compatibility with FPGA implementation, we impose the following constraints during training:
\begin{itemize}
    \item All weights and biases are clipped to lie within the range $[-4, 4]$.
    \item The decimal precision of the parameters is limited to four digits after the decimal point to match the fixed-point representation supported by the hardware.
\end{itemize}

After training, all network parameters (weights and biases) are quantized and converted into fixed-point format suitable for deployment. We use 18-bit representation for weights (6 bits integer, 12 bits fractional) and 27-bit representation for inputs and intermediate activations (10 bits integer, 17 bits fractional), ensuring numerical stability and efficient use of DSP resources on the FPGA.

The small parameter count and constrained precision enable real-time inference with minimal latency, making the network well-suited for mid-circuit measurements and other timing-critical applications in quantum computing.

For our evaluation, we utilized the AMD Xilinx Zynq UltraScale+ RFSoC ZCU216 FPGA \cite{amd2024zcu216}, integrating our system. Our primary objective was to ensure that the hardware implementation achieves performance comparable to its software-based counterpart, with the added advantage of real-time processing capabilities.

The integrated readout data, post-demodulation, undergoes scaling and normalization, which is critical before feeding data into a neural network to ensure consistent input scaling, which improves model stability and generalization \cite{fpga_sat, normalization}. However, on FPGAs, traditional methods such as Z-score or linear scaling are challenging due to expensive division operations. To address this, we implement a shift-based linear scaling approach by leveraging the FPGA’s efficiency in right-shift operations for division by powers of two.

\textbf{Mean Shifting.}  
The mean of both I and Q components is shifted to [0, 0], centering the distribution across all four Cartesian quadrants for balanced normalization (Fig ~\ref{norm_mov}).

\textbf{Shift-Based Normalization.}  
Given symmetric max and min values (\(|\text{max}| \approx |\text{min}|\)), the normalization is expressed as:
\begin{align}
    \text{Norm} &= \frac{X - \text{min}}{\text{max} - \text{min}} \\[6pt]
    &\approx \frac{X + 2^n}{2^{n+1}} \\[6pt]
    & = (X + 2^n) \gg (n+1)
    \label{eq: norm_modi1}
\end{align}

This allows implementation using only shift operations.

\begin{algorithm}[t]
\small
\caption{FPGA-Friendly Data Scaling}
\begin{algorithmic}[1]
\State $\text{data} \gets \text{input scalar}$
\State $n \gets \text{nearest } 2^n \text{ to max value}$
\State $\mu \gets \text{mean of data}$
\State $\text{tmp} \gets \text{data} + 2^n - \mu$
\State $\text{data}_{\text{norm}} \gets (\text{tmp} \ll 17) \gg (n+1)$
\end{algorithmic}
\label{algo: Norm}
\end{algorithm}

\textbf{Decimal Representation.}  
To preserve fractional values, we use the following method to avoid division by 10,000:
\begin{align*}
(a) &\quad val = (X + 2^{n}) \ll 14 \\
(b) &\quad tmp = val \gg (n+1) \\
(c) &\quad data_{norm} = (tmp \ll 17) \gg 14
\end{align*}
This provides a fixed-point representation (10 integer bits, 17 fractional bits) used in the neural network. Combining ($a$),(b), and (c), the final normalization procedure can be obtained. Algorithm \ref{algo: Norm} shows the entire data scaling procedure, including mean shifting and normalization. This algorithm is applied to I and Q with different values for $n$, where $n$ is precomputed during training. It's important to note that we use this same data scaling procedure while training to avoid train-test discrepancies.

\subsection{\textbf{FPGA implementation}}

These normalized IQ values are then fed into our hardware-based neural network architecture. We adopted a modular approach to implement the feedforward neural network (FNN) architecture, segmenting the network into sequences of modules. Each layer, except the input layer, is followed by an activation module and comprises a list of nodes (neurons). As depicted in Fig \ref{fig:fpga_implementation}, each node primarily performs multiplication followed by an addition operation. The multiplication operation involves the weights of the current layer and the output from the previous activation module or input layer.

In our architecture, weights are fixed at 18 bits, and inputs (outputs from the previous layer) are 27 bits. Both operands are fed into DSP48 slices, producing a 45-bit output (16-bit integer part and 29-bit fractional part). We extract a 27-bit value from this output by selecting the 10 least significant bits (LSBs) from the integer part and the 17 most significant bits (MSBs) from the fractional part for further processing. The outputs of these DSP slices are summed along with a 27-bit bias, as represented by the equation:

\begin{equation} W_{0}I + W_{1}Q + B \end{equation}

In our FPGA-based architecture, each multiplication operation requires two clock cycles (4ns), while addition operations are executed within a single clock cycle (2ns). Typically, additions involve two operands sourced from DSP outputs; however, certain scenarios necessitate three operands, including two DSP outputs and one bias term. To optimize processing efficiency, we constrain summations to a maximum of three operands within a single clock cycle. For instances where more than three operands need to be added, we employ a sequential approach, distributing the summation across that layer while maintaining parallelism across the nodes of that layer. For example, Layer 2 contains nine operands (e.g., eight DSP outputs and one bias term) within a single node; thus, a sequential summation strategy is employed due to the impracticality of executing this operation within one clock cycle. Layer 2 operates with three such nodes, contributing to a total latency of six clock cycles.

Layer 1 in our FPGA architecture is followed by a ReLU (Rectified Linear Unit) module. This module determines whether the input from a node is positive or negative: if positive, the output remains unchanged; if negative, the output is set to zero. The ReLU module operates within one clock cycle.

After the final layer, we implement a softmax function to obtain probability distributions over the output classes. The softmax implementation consists of two lookup table (LUT) operations: one for the exponential term and another for the inverse term. We first look at the exponential function $e^{x}$ for all three states, where $x$ is the output from the last layer with three nodes (representing a three-state system). Additionally, we compute:

\begin{equation}
    D = \frac{1}{e^{x_{0}} + e^{x_{1}} + e^{x_{2}}} 
    \label{eq: sigmoid denominator} 
\end{equation}

Finally, we multiply each $e^{x}$ by $D$ to obtain the softmax probabilities. The final step involves determining the maximum value among the three probabilities. The entire process from normalization to softmax takes 33 clock cycles. By omitting the softmax computation and directly determining the maximum from the last layer's outputs, the final result can be obtained in 20 clock cycles.

\subsection{\textbf{Data Availability}}
All data relevant to the study are available from the corresponding authors upon request.

{\section{\textbf{Acknowledgment}}}
\label{sec:ack}
This material is based upon work supported by the U.S. Department of Energy, Office of Science, National Quantum Information Science Research Centers, Quantum Systems Accelerator (Award No. DESCL0000121). Additional support is acknowledged from the U.S. Department of Energy, Office of Science, the Office of High Energy Physics, Advanced Scientific Computing Research Testbeds for Science program under Contract No. DE-AC02-05CH11231, the National Science Foundation under Award Numbers 2152357 and 2401415, and the Defense Advanced Research Projects Agency under Grant Number HR0011-24-9-0358.

Y.X. and N.R.V. conceived and designed the system. N.R.V. and Y.X. carried out the FPGA implementation and performed the experiments. A.H. and N.G. proposed the validation experiments on superconducting qubits. H.N.N. and H.L. provided fruitful discussions on the machine learning model design. G.H., Y.X., and N.F. developed the QubiC system used in this work. J.B., A.R., and K.N. provided assistance. Q.J., K.B.W., I.S., G.H., and P.N. secured funding and provided resources. N.R.V. wrote the manuscript, and Y.X. and P.N. reviewed and edited it. Y.X. and P.N. supervised the project.

Y.X. has a financial interest in OpenQI, Inc.
\bibliography{reference}
\bibliographystyle{unsrt}

\end{document}